\documentclass[fleqn,usenatbib]{mnras}

\usepackage{newtxtext,newtxmath}

\usepackage{amsmath}

\usepackage{multirow}
\usepackage{makecell}
\usepackage{comment}

\usepackage[T1]{fontenc}
\usepackage{array}
\DeclareRobustCommand{\VAN}[3]{#2}
\let\VANthebibliography\thebibliography
\def\thebibliography{\DeclareRobustCommand{\VAN}[3]{##3}\VANthebibliography}

\usepackage{graphicx}	
\usepackage{amsmath}	

\usepackage{url}
\usepackage{hyperref}

\usepackage[usenames,dvipsnames]{xcolor}

\usepackage{float}

\title[Probing cosmic isotropy in the Local Universe]{Probing cosmic isotropy in the Local Universe}

\author[C. Franco et al.]{
	Camila Franco,$^{1}$\thanks{E-mail: camilafranco@on.br}
	Felipe Avila,$^{1}$ 
    Armando Bernui$^{1}$ \\
	$^{1}$Observat\'orio Nacional, Rua General Jos\'e Cristino 77, 
	S\~ao Crist\'ov\~ao, 20921-400 Rio de Janeiro, RJ, Brazil \\
}

\date{Accepted 2023 November 20. Received 2023 November 20; in original form 2023 May 26}

\pubyear{2023}

\begin{document}
\label{firstpage}
\pagerange{\pageref{firstpage}--\pageref{lastpage}}
\maketitle

\begin{abstract}
This is a model-independent analysis that investigates the statistical isotropy in the Local Universe using the ALFALFA survey data
($0 < z < 0.06$). We investigate the angular distribution of HI extra-galactic sources from the ALFALFA catalogue and study whether they are compatible with the statistical isotropy hypothesis using the two-point angular correlation function (2PACF). Aware that the Local Universe is plenty of clustered structures and large voids, we compute the 2PACF with the Landy-Szalay estimator performing directional analyses to inspect 10 sky regions. We investigate these 2PACF using power-law best-fit analyses, and determine the statistical significance of the best-fit parameters for the 10 ALFALFA regions by comparison with the ones obtained through the same procedure applied to a set of mock catalogues produced under the homogeneity and isotropy hypotheses. Our conclusion is that the Local Universe, as mapped by the HI sources of the ALFALFA survey, is in agreement with the hypothesis of statistical isotropy within $2\,\sigma$ confidence level, for small and large angle analyses, with the only exception of one region --located near the Dipole Repeller-- which appears slightly outlier ($2.4\,\sigma$).
Interestingly, regarding the large angular distribution of the HI sources, we found 3 regions where the presence of cosmic voids reported in the literature left their signature in our 2PACF, suggesting projected large underdensities there, with number-density contrast $\delta \simeq -0.7$.
According to the current literature these regions correspond, partially, to the sky position of the void structures known as Local Cosmic Void and Dipole Repeller.
\end{abstract}


\section{Introduction}\label{sec:intro}

At the end of the last century it was observed the presence of an  intriguing large underdense region in the neighborhood of the Milky Way~\citep{Tully87}, a region with low density of luminous 
matter inside it, currently termed the {\em Local Cosmic Void}~\citep[LCV;][]{Tully08,Tully13,Tully19}.

The LCV has been the target of detailed observations to determine its main features: location, size, shape, etc. 
Recent observational efforts provide more information of the nearby universe and confirm the presence and vastness of a large void in the backyard of the Milky Way. 
These observations indicate that the LCV is a 150 -- 300 Mpc void~\citep{Plionis02,Keenan,Whitbourn14,Whitbourn16,Tully08,Tully13,Tully19,Bohringer20}. 
The presence of the LCV near the Milky Way attracts the attention due to its, apparently, huge size~\citep{Keenan,Tully19}, but it is also the focus of studies for other reasons. 
One important motivation for studying the LCV is due to the dynamical effect that it causes, producing large peculiar velocities in the nearby 
galaxies~\citep{Tully08,Tully13,Tully19,Hoffman17}, 
information that is needed, for example, to a better calibration of the standard candles that helps to measure $H_0$.

Cosmological simulations, based on the $\Lambda$CDM concordance model, predict that the universe is a cosmic web plenty of filaments, with clustered matter at their intersections, and voids of different sizes scattered everywhere~\citep{Springel06, Pandey11,Vogelsberger14, Sarkar22}. 
Such a large dimension reported for the LCV may defy the homogeneity and isotropy features expected in the $\Lambda$CDM model~\citep{Peebles10}. 
However, it is also possible that medium-sized voids located next to each other give the impression of a very large void, as observed in the distribution of voids found in the Local Universe~\cite[][see Figure 1 in this 
reference]{Moorman14}.
In fact, in the same way that filaments connect ones to the others, the voids network may interconnect small and medium size voids that can be observationally interpreted as larger voids. We shall explore this possibility in our analyses.

The Cosmological Principle (CP), fundamental part 
of the standard model of cosmology, has been tested using data from various astronomical surveys with different cosmic tracers~\citep{Bolejko09,Labini10,Aluri22,Avila22}, such as quasars~\citep{Secrest21,Goncalves18b,Fujii22}, CMB~\citep{Aluri12,Marques18,Khan22,Khan22b}, gravitational waves~\citep{Galloni22}, gamma-ray bursts~\citep{Bernui08,Jakub2019}, 
galaxy clusters~\citep{Bengaly17}, and 
galaxies~\citep{Appleby14,Goncalves18,Avila19}, between others.

The application of the CP to the Local Universe is an interesting issue {\it di per se} due to our ignorance of all the structures present and how they are clustered, in particular because many of these structures are hidden to us by the Milky Way disc~\citep{Guaianazzi05,Hwang13}. 
In this respect, the LCV is also a fascinating piece in the puzzle of how matter and void structures are distributed in the Local Universe~\citep{Whitbourn14,Hoffman17,Tully19}.

We investigate the angular distribution of the HI extragalactic sources catalogued by the ALFALFA survey, analysing the two-point angular correlation function (2PACF) in a set of sky patches in which we divide the survey footprint, and then compare the result with random data sets and mock synthetic catalogues, both constructed under the hypothesis of statistical homogeneity and isotropy. 
Our methodology is model-independent, as the 2PACF is computed only with information on the angular position of the objects. 
Additionally, the errors in the 2PACF are determined using the covariance matrix, which uses lognormal catalogues produced assuming a fiducial cosmology (see section~\ref{log-normal}).

Some words are in due to explain the motivations to 
probe the cosmic isotropy in the Local Universe analysing the HI extragalactic sources of the ALFALFA survey. 
These data present suitable features for the scope of our analyses, 
where these features include: 
(i) a good number density of cosmic objects, 
$n \sim 4\,\text{deg}^{-2}$; 
(ii) a large surveyed sky area, almost $1/6$ of the celestial sphere, necessary to divide it into many patches; 
(iii) the relative bias of the HI sources, with respect to matter, is  close to the bias of the blue galaxies~\citep{Papastergis13,Avila18}, which in turn is close to 1~\citep{Cresswell09,Avila19}, therefore HI appear to be a reasonable cosmic tracer of matter in the Local Universe\footnote{Not to mention that the catalogued objects were detected with SNR $> 6.5$, with confirmed optical counterpart.}.

The outline of this work is the following. 
In Section~\ref{sec:data} we present the details of the ALFALFA data we use in 
our scrutiny of the Local Universe. 
In Section~\ref{sec:method} we describe the 2PACF, the mock catalogues and the 
estimation of uncertainties. 
In Section~\ref{sec:results} we present our analyses and discuss our results; 
we leave for Section~\ref{sec:final} our final conclusions. 
In the appendix~\ref{ap:simulated-void} we discuss the signature left by a simulated void; finally, in appendix~\ref{ap:tests} we present a null test that shows the consistency of our procedures.

\section{The Arecibo Legacy Fast ALFA Survey} \label{sec:data}

The Arecibo Legacy Fast ALFA Survey\footnote{\url{http://egg.astro.cornell.edu/alfalfa/data/index.php}} (ALFALFA) is a blind $21$~cm HI emission-line catalogue,
which used the seven-horn Arecibo L-band Feed Array (ALFA) to cover an area of $ \sim 7000 \,\text{deg}^{2}$ with the main objective of obtaining a robust determination of the faint end of the HI mass function (HIMF)~\citep{Haynes18,Papastergis13,Giovanelli05}. Most of the sources detected are gas-rich galaxies with low surface brightness and dwarfs that populate the Local Universe $0 < z < 0.06$~\citep{Giovanelli15,Haynes18}. 
Recently, it has been released the final version of the ALFALFA catalogue, with 100\% of the planned footprint area, containing $31502$ extragalactic sources classified in two categories according to its HI line detection status: {\sc code} 1, high signal to noise ratio (SNR $> 6.5$) extragalactic sources, considered highly reliable and with confirmed optical counterpart; and {\sc code} 2, lower signal to noise ratio (SNR $\lesssim 6.5$) HI signal coincident with optical counterpart at the same position and redshift, considered unreliable sources~\citep{Haynes18,Martin12}\footnote{For other applications of the ALFALFA catalogue see, e.g.,~\cite{Papastergis13,Avila21}}.

\begin{figure}
\begin{minipage}[b]{\linewidth}
\includegraphics[width=1.02\textwidth]{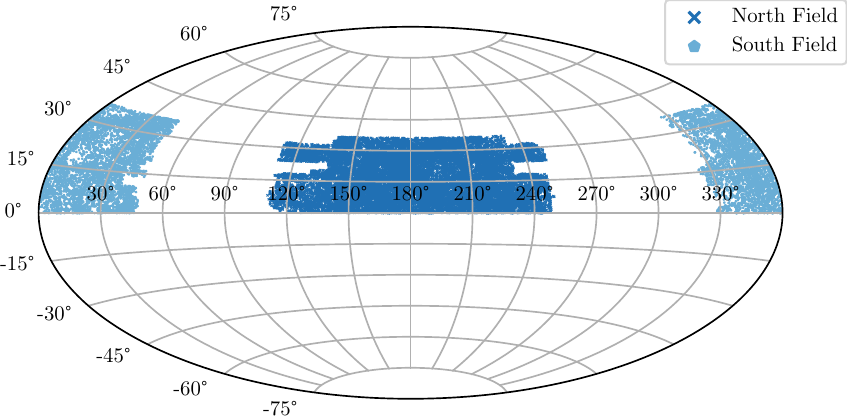}
\end{minipage}
\caption{The ALFALFA footprint in equatorial coordinates. 
Notice that the contour is not completely regular.}
\label{fig:alfalfa-footprint}
\end{figure}

In this work we shall use only the sources designated as {\sc code~1}, as recommended by the ALFALFA team. The ALFALFA survey covers two regions, both in the declination range
$0^{\circ} < \text{DEC} < 36^{\circ}$, and in the right ascensions intervals of $21^{\text{h}} 30^{\text{m}} < \text{RA} < 3^{\text{h}} 15^{\text{m}}$ and $7^{\text{h}} 20^{\text{m}} < \text{RA} < 16^{\text{h}}40^{\text{m}}$ (see Figure~\ref{fig:alfalfa-footprint}). The first region, referred as {\it Fall region} (in the Southern Galactic Hemisphere, SGH), contains a total of $9924$ HI sources and the second one, the {\it Spring region} (in the Northern Galactic Hemisphere, NGH), have $21578$ HI sources.

\section{Methodology}\label{sec:method}

In this section we explain the methodology that we follow to study the angular distribution of the ALFALFA cosmic objects by applying the 2PACF. 
Notice that these analyses are model-independent because this function only uses the angular position of the ALFALFA data, 
plus random catalogues generated considering random angular positions in the data footprint (produced without assuming any 
cosmological model). 

\subsection{Two-Point Angular Correlation Function (2PACF)}

To perform our clustering analyses of the ALFALFA extragalactic HI sources, projected onto the celestial sphere, to calculate the 2PACF we adopt the Landy-Szalay estimator (LS; \cite{LS93}) 
\begin{equation}\label{eq:omega-ls}
\omega_{LS}(\theta) \equiv \frac{DD(\theta) - 2DR(\theta) 
+ RR(\theta)}{RR(\theta)} \,,
\end{equation}
where $DD(\theta)$ is the number of galaxy pairs in the sample data with angular separation $\theta$, normalized by the total number of pairs; $RR(\theta)$ is a similar quantity, but for the pairs in a random sample; and $DR(\theta)$ corresponds to a cross-correlation between a data object and a random object. 
Our model-independent analyses only use the measured angular coordinates of the object sky position to calculate the angular distance between pairs, given by
\begin{equation}\label{eq:theta}
    \theta_{ij} = \cos^{-1}[\sin(\delta_i)\sin(\delta_j) + \cos(\delta_i)\cos(\delta_j)\cos(\alpha_i - \alpha_j)] \,,
\end{equation}
where $\alpha_i$, $\alpha_j$ and $\delta_i$, $\delta_j$ are the right ascension and the declination, respectively, of the galaxies $i$ and $j$.

To calculate the 2PACF, one needs a number of random sets $N$, that is, isotropic distributions of point sources with the same observational features as the data set in analysis: the same footprint and equal number of objects as the sample in study. The construction of the random sets followed the description presented by~\cite{Bernui04} (see also~\cite{deCarvalho18, Keihanen19, Wang13}).  

After numerical evaluations, we consider $N=25$ random sets for each region under analysis, because for $N > 25$ the behaviour of the 2PACF is the same.

According to the hypotheses of homogeneity and isotropy, basis of the concordance model, the expectation regarding the 2PACF is that it behaves according to a power-law~\citep{Peebles93,Coil13,Kurki-Suonio,Connolly02,Marques20,Totsuji69}, like 
\begin{equation}\label{eq:omega}
\omega (\theta) = \left(\frac{\theta}{\theta_0} \right)^{-\beta}\,, 
\end{equation}
where the parameter $\theta_0$ represents the transition scale between the non-linear and linear regimes, and the parameter $\beta$ quantifies the matter clustering (i.e., the higher the value of this parameter, the more galaxies are close to each other (see, e.g.,~\cite{Peebles93, Coil12,Connolly02,Marques20,Marques23}).

\subsection{Lognormal random field distributions}\label{log-normal}

In the last decade the literature has reported that the error computed in the 
two-point correlation analyses is underestimated using random catalogues~\citep{Norberg09,deCarvalho18}. 
An error estimate close to the real situation can be achieved with the lognormal 
catalogues~\citep{Norberg09,deCarvalho20}. 
In this section, we shall briefly describe the main features of the lognormal random field distributions.

The current scenario is to assume a Gaussian random field to describe the 
matter-density contrast field\,\footnote{As usual, the matter-density contrast is defined by $\delta \equiv \frac{\delta \rho}{\overline{\rho\,}}$.}, $\delta(\textbf{x})$, because 
(i) it is predicted by most inflationary models~(see \cite{Barrow90} and references therein), 
and (ii) the model is fully specified, i.e., statistically complete~\citep{Coles91}\footnote{Gaussian random fields are completely determined once we specify the mean and the covariance~\citep{Adler10}.}. 
Thus we need only the two-point correlation function $\xi(r)$, or its Fourier transform $P(k)$, to describe the matter density  field~\citep{Agrawal17}. 

However, Gaussian random fields cannot describe the matter distribution, it assigns non-zero probability for negative densities~\citep{Fry86}. 
Although small in the beginning, the matter fluctuations grow as gravitational instability takes over, thus it cannot describe consistently the non-linear density matter distribution at late times. 
This motivates the construction of \textit{stochastic models} for the density field: models which are completely specified statistically, but do not violate $\overline{\rho} > 0$~\citep{Coles91}. 

One model that happen to fulfill these two conditions is the lognormal random field~\citep{Coles91,Colombi94,Kofman94,Bernardeau95,Uhlemann16,Shin17}. 
This particular field can be obtained by transforming a Gaussian field $X(\textbf{r})$ as
\begin{equation}
Y(\textbf{r}) = \exp[X(\textbf{r})]\,,
\end{equation}
resulting in the distribution of $Y$~\citep{Coles91} 
\begin{equation}
f(y) = \frac{1}{\sigma\sqrt{2\pi}}\exp\left[-\frac{(\log y-\mu)^2}{2\sigma^2}\right]\frac{dy}{y}\,,
\end{equation}
where $\mu$ and $\sigma^2$ are the  mean and variance of the Gaussian field, respectively. 
To specify all the statistical properties we extend the formalism for a multivariate normal distribution 
\begin{eqnarray}
\hspace{-0.5cm}
f_n(y_1,...,y_n) \!\!\!\!&=& \!\!\!\!(2\pi)^{-n/2}|\textbf{M}|^{-1/2} \nonumber\\
&\times& \!\!\!\!\exp\left[-\frac{1}{2}\sum_{i,j}\textbf{M}_{ij}^{-1}\log(y_i)\log(y_j)\right]\Pi_{i=1}^{n}\frac{1}{y_i} ,
\end{eqnarray}
where \textbf{M} is the covariance matrix defined as
\begin{equation}
\textbf{M} \equiv \langle(X_i-\mu)(X_j-\mu)\rangle \,.
\end{equation}

Considering the advantages and disadvantages of describing the density field assuming a lognormal random field, nowadays we have a variety of public codes to generate simulated galaxy catalogues~\citep{Marulli16,Xavier16,Agrawal17,Hand18,Ramirez22}. For each public code the accompanying paper presents validation tests comparing inputs and results. 
Although N-body simulations remain the main tool for describing no-linearity of the galaxy distribution, the use of N-body simulations to constrain cosmological parameters demand too large computing resources~\citep{Agrawal17}. 
Using a lognormal code we can generate thousand of mocks to compute a robust covariance matrix. 
In addition, recent works show that for the main statistical estimators of the galaxy distribution, namely correlation function, power spectrum, and bispectrum, the N-body simulations and the lognormal mocks present similar covariance matrix results~\citep{Lippich19,Blot19,Colavincenzo19}.

For this work we use the public code of \cite{Agrawal17}. In a previous work we used this code to obtain the covariance matrix for the dipole analysis in the Local Universe with the ALFALFA catalogue~\citep{Avila21}. As \cite{Agrawal17} comments, the code is fast, direct and instructive. Also, for our purpose, the principal issues in the non-linear scales do not affect our results because we are interested in the large angular scales. See \cite{Agrawal17} 
for details on generating lognormal catalogues.

The input parameters needed to generate the mock catalogues that reproduce the Local Universe clustering features are listed in Table~\ref{tab:survey_config}. 
They are: the redshift $z$ (the mean and median redshift of the catalogue is 
$z \simeq 0.026$, we choose $z = 0$), the bias $b$ (see~\cite{Avila21} for more information), the number of galaxies $N_g$, and the box dimensions\footnote{In units of \,Mpc$/h$.} ($L_x, L_y, L_z$). 
We set the mesh number as 
$\text{Pnmax} = 128$\footnote{We also tested the catalogues with $\text{Pnmax} = 512$ and similar results were obtained.}. 
The code also asks for cosmological parameters. 
In Table~\ref{tab:cosmo_param} we show the cosmological parameters from the~\cite{Planck20}, used to produce these lognormal catalogues.

\begin{table}
\centering
\caption{Survey configuration used to generate the set of $1000$ lognormal mock catalogues used in our analyses. $N_g$ is the number of galaxies in the box configuration, with dimensions ($L_x,L_y,L_z$).}
\scalebox{1.15}{
\begin{tabular}[t]{lcc}
\hline
&Survey configuration&\\
\hline
&$z = 0.0$\\
&$b = 1.0$\\
&$N_g = 2 \times 10^5$\\
&$L_x = 230$ Mpc$/h$ \\
&$L_y = 230$ Mpc$/h$ \\
&$L_z = 230$ Mpc$/h$ \\
\hline
\end{tabular}
}
\label{tab:survey_config}
\end{table}

\begin{table}
\centering
\caption{Cosmological parameters from the Planck Collaboration last data release~\citep{Planck20}.}
\scalebox{1.15}{
\begin{tabular}[t]{lcc}
\hline
&Cosmological parameters&\\
\hline
&$\Omega_b h^2 = 0.02236$\\
&$\Omega_c h^2 = 0.1202$\\
&$\ln(10A_s) = 3.045$\\
&$n_s = 0.9649$\\
&$\Sigma m_{\nu} = 0.06$ eV\\
&$h = 0.6727$\\
\hline
\end{tabular}
}
\label{tab:cosmo_param}
\end{table}

\vspace{0.5cm}
\subsection{Covariance matrix estimation}

With the mock catalogues, one can calculate the covariance matrix in order to obtain realistic  uncertainties in the 2PACF. The procedure is the following (see, e.g.,~\cite{deCarvalho21,Myers05,Wang13}). 
For each mock, the 2PACF was calculated from a set of $N_b$ bins and the matrix was estimated according to  
\begin{equation}\label{eq:cov}
    \text{Cov}_{ij} = \frac{1}{N} \sum_{k = 1}^{N}[\omega_k(\theta_i) - \overline{\omega}_k(\theta_i)][\omega_k(\theta_j) - \overline{\omega}_k(\theta_j)],
\end{equation}
where the indices $i, j = 1,2,...,N_b$ represent each bin $\theta_i$; $\omega_k$ is the 2PACF for the $k$-th mock ($k = 1,2,...,N$); $\overline{\omega}(\theta_i)$ and $\overline{\omega}(\theta_j)$ are the mean value at the bin $i$ and $j$, respectively. 
The uncertainty in the 2PACF plots is the square root of the main diagonal of Equation~(\ref{eq:cov}), $\Delta{\omega(\theta_i)} = \sqrt{\,\text{Cov}_{ii}}$, but the likelihood analyses (which includes the next best-fits evaluation) is obtained using the full covariance matrix, which takes into account the full correlation matrix (with all the bins).

In order to verify how uncorrelated the bins are with each other, we calculate the terms of the correlation matrix using the relation
\begin{equation}\label{eq:corr}
\text{Corr}_{ij} = \frac{\text{Cov}_{ij}}{\sigma_{i}\sigma_{j}}\,,
\end{equation}
where $\sigma_{i}, \sigma_{j}$ are the standard deviations of the bins $i$, $j$, respectively~\citep{Wasserman04}. 

\begin{figure}
\begin{minipage}[b]{\linewidth}
\centering
\includegraphics[width=\textwidth]{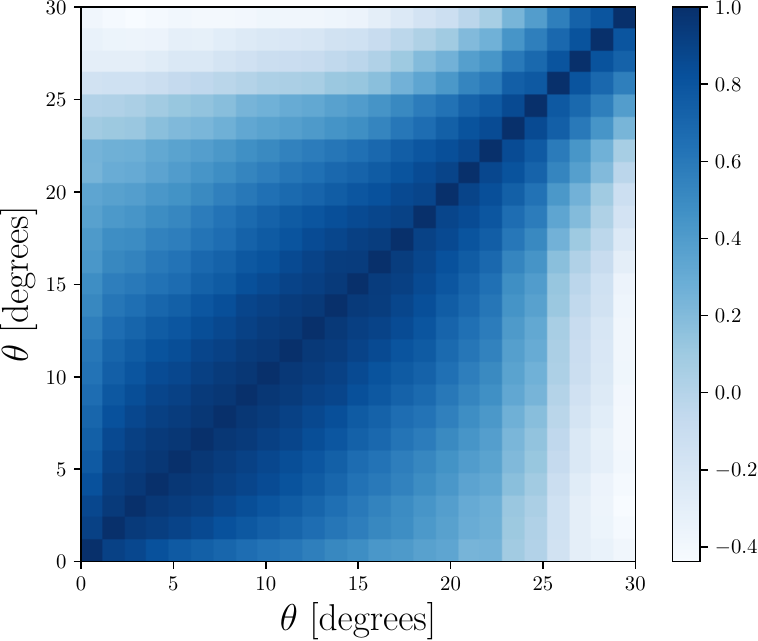}
\end{minipage}
\caption{Correlation matrix for 100 lognormal mocks obtained from Equation~(\ref{eq:corr}). 
Our 2PACF analyses consider $\theta = 30^{\circ}$ as the maximum angular separation (see section~\ref{selectedregions} for details).}
\label{fig:correlation-matrix}
\end{figure}

In Figure~\ref{fig:correlation-matrix} we show the correlation matrix for 100 lognormal mocks.
It is worth noting that the color scale indicates whether the bins are independent ($\text{Corr}_{ij} = 0$) or whether increasing bin $i$ causes an increase or decrease in bin $j$ ($\text{Corr}_{ij} = +1$ or $\text{Corr}_{ij} = -1$, respectively).

In the next section we shall perform the minimization procedure, 
where we chose the best-fit parameters $\theta_0$ and $\beta$ 
that minimize the $\chi^{2}$ function 
\begin{flalign}
&\chi^{2}(\theta; \beta, \theta_0)& \nonumber \\
&=[\omega(\theta) - \omega_{\text{model}}(\theta, \beta, \theta_0)]^{T} \text{Cov}^{-1}[\omega(\theta) - \omega_{\text{model}}(\theta, \beta, \theta_0)]\,,&
\end{flalign}
where $\text{Cov}^{-1}$ is the inverse of the covariance matrix.

\section{Analyses and Results with the 2PACF}\label{sec:results}

With the statistical tools on hands described in the previous section, we perform the statistical isotropy analyses on the ALFALFA data. 
Firstly, we perform the 2PACF analyses for our sample --carefully fragmented in 10 directions-- to probe the Local Universe isotropy. 
Next, we discuss the interesting features observed in these 2PACF, like the possible presence of \textit{low-density} structures in the analysed regions. 
We consider small and large angular analyses in the fitting procedure 
of the 2PACF; the first one because of what is established in the 
literature~\citep{Wang13}, while the second is due to the intriguing 
structures observed only in some regions at large angles.

\subsubsection{The selected 10 regions for 2PACF analyses}
\label{selectedregions}

\begin{figure*}
\begin{minipage}[b]{\linewidth}
\includegraphics[width=0.49\textwidth]{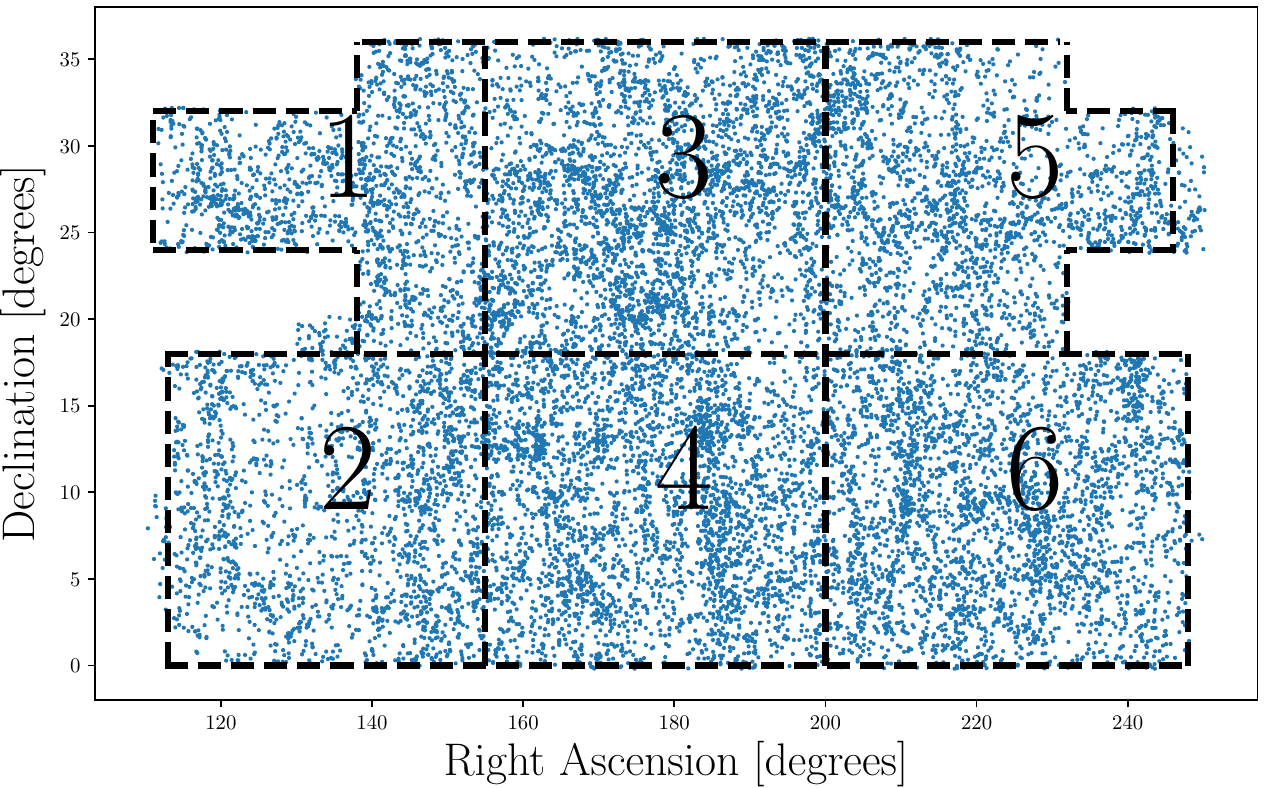}
\includegraphics[width=0.49\textwidth]{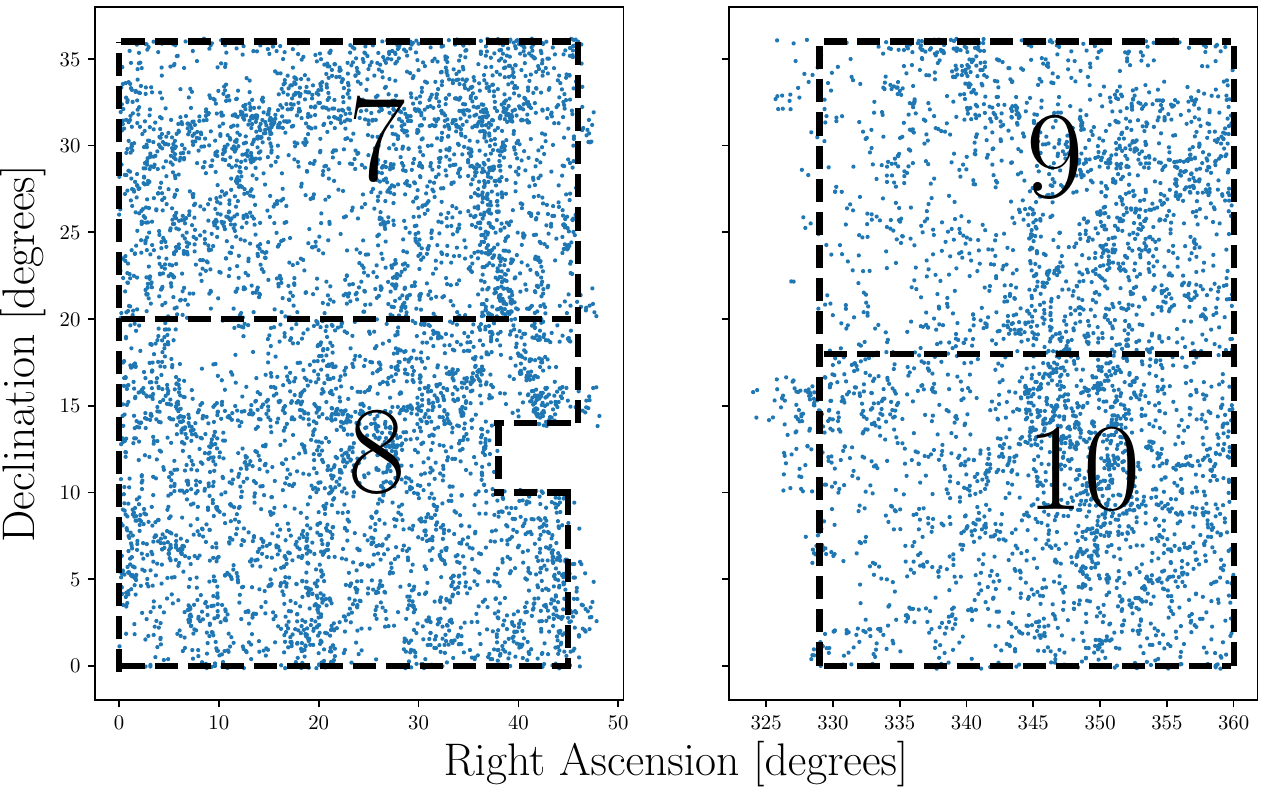}
\end{minipage}
\caption{ALFALFA footprint division in 10 sky patches for statistical isotropy analyses using the 2PACF. 
The regions 1--6 and 7--10 correspond to the Spring (NGH) and Fall (SGH) regions, respectively. 
Notice that some small cuts of data were done in order to smooth the areas for analyses.}
\label{fig:alfalfa-division}
\end{figure*}

Our data sample, suitable for statistical isotropy analyses, has $16285$ 
extragalactic sources in the NGH and $9149$ extragalactic sources in the SGH, 
distributed in $\sim \!4000\,\text{deg}^{2}$ and $\sim \!3000\,\text{deg}^{2}$, 
respectively. 
Considering these data, we shall perform directional analysis dividing the 
survey footprint in a number of patches. 
The suitable size for the construction of the polygons is a delicate issue. 
To minimize boundary effects, due to irregular contours of the survey geometry, 
the patches for analyses are not necessarily quadrilaterals but polygons. 
Another important reason for the polygons not be so small is the number density of cosmic objects; in fact our tessellation produced regions with a similar number density to avoid a large statistical noise, and approximately similar  area $\sim 700\,\text{deg}^{2}$. 
The area of these sky patches can not be so small because the statistical noise shall dominate the 2PACF analysis there, and it can not be so big because we intend to study a large number of directions. 
Also, to avoid irregularities at the edges and facilitate the reproduction of the footprints in the simulated catalogues, some cosmic objects had to be discarded avoiding polygons with many sides. 
According to these criteria, the final sample for analyses has $15993$ and $8828$ extragalactic sources in the NGH and SGH, respectively. 
Then, we arrived to the compromise of considering regions of similar area, 
shown in the Figure~\ref{fig:alfalfa-division}, with a mean number density $n \simeq 3.4\,\text{deg}^{-2}$. 

After some tests we select for analyses 10 directions. 
For this we divide the ALFALFA footprint into 10 sky patches of $\sim \!700\,\text{deg}^{2}$ each one, 
6 in the Spring region and 4 in the Fall region, as shown in Figure~\ref{fig:alfalfa-division}, 
with geometric details specified in Table~\ref{tab:areas}. 
Using the LS estimator, we then calculate the 2PACF, $\omega(\theta)$, for the range 
$0^{\circ} < \theta < 40^{\circ}$, with $80$ bins of width $\Delta \theta = 0.5^{\circ}$, in each sky patch. 
However, performing the 2PACF analyses in the selected regions we observed that for pairs separations larger than $\theta \gtrsim 30^{\circ}$ the number of pairs decreases substantially, in consequence the 2PACF is no longer informative of the matter clustering we are trying to describe. 
For this reason, our 2PACF analyses consider $\theta = 30^{\circ}$ as the maximum angular separation.

We shall perform a best-fit procedure assuming a power-law behavior in the form of Equation~(\ref{eq:omega}), considering two situations: 
(i) for small angular scales 
$0^{\circ} < \theta < 10^{\circ}$; and 
(ii) for large angular scales 
$0^{\circ} < \theta < 30^{\circ}$.

\begin{table}
\centering
\caption{Features of the 10 regions under scrutiny. 
Observe that, despite our efforts, the regions still show some (small) differences in the assigned area. But the 
density number parameter, $n$, is less scattered. The uncertainties were calculated using $\sigma_{n} = \sqrt{N}/\text{area}$, where $N$ is the number of sources.}
\scalebox{1.15}{
\begin{tabular}[ht]{lccc}
\hline
&\hspace{0.cm} sources &\hspace{0.cm} area [$\text{deg}^{2}$] &\hspace{0.cm} $n\,[\text{deg}^{-2}]$\\
\hline
Area 1 & $1648$&$522$&$3.15\,\pm\,0.08$\\
Area 2 & $2081$&$756$&$2.75\,\pm\,0.06$\\
Area 3 & $3372$&$810$&$4.16\,\pm\,0.07$\\
Area 4 & $3427$&$810$&$4.23\,\pm\,0.07$\\
Area 5 & $2157$&$688$&$3.13\,\pm\,0.07$\\
Area 6 & $3308$&$864$&$3.83\,\pm\,0.07$\\
Area 7 & $2474$&$736$&$3.36\,\pm\,0.07$\\
Area 8 & $2844$&$875$&$3.25\,\pm\,0.06$\\
Area 9 & $1636$&$558$&$2.93\,\pm\,0.07$\\
Area 10 & $1874$&$558$&$3.36\,\pm\,0.08$\\
\hline
\end{tabular}
}
\label{tab:areas}
\end{table}

\subsection{2PACF small-angle analyses}
The best-fit analyses for small angles, $\theta_0^{S}$ and $\beta^{S}$, deserve a comment. 
The analysis for small angles, $0^{\circ} < \theta < 10^{\circ}$, provides a measure of non-linear clustering 
in 10 sky directions that we compare with $1000$ {\bf Area}-mocks\footnote{That is, 100 mocks where we consider 
in each one 10 regions with equal footprint as in Figure~\ref{fig:alfalfa-division}.} produced under the homogeneity and isotropy hypotheses. 
This comparison is intended to reveal possible deviations of the best-fit parameters from the 10 ALFALFA regions 
with respect to the isotropic mocks, simulated data that take into account the clustering evolution of the low-redshift universe. 
For small-angle analyses, the median and standard deviation of the best-fit parameter $\beta^{S}$ from these $1000$ {\bf Area}-mocks are \,$\beta^{S;\,SI} = 1.180 \pm 0.325$. 
For the small-angle analyses the histogram of the $\beta^{S}$ values obtained from the $1000$ {\bf Area}-mocks analysed is displayed in the left panel of Figure~\ref{fig:beta-dist}. 
We conclude that the 10 ALFALFA regions investigated are compatible with the hypothesis of statistical isotropy within $2\,\sigma$ confidence level (CL).

\begin{figure*}
\begin{minipage}{\linewidth}
\centering
\includegraphics[width=15.0cm,height=22.5cm]{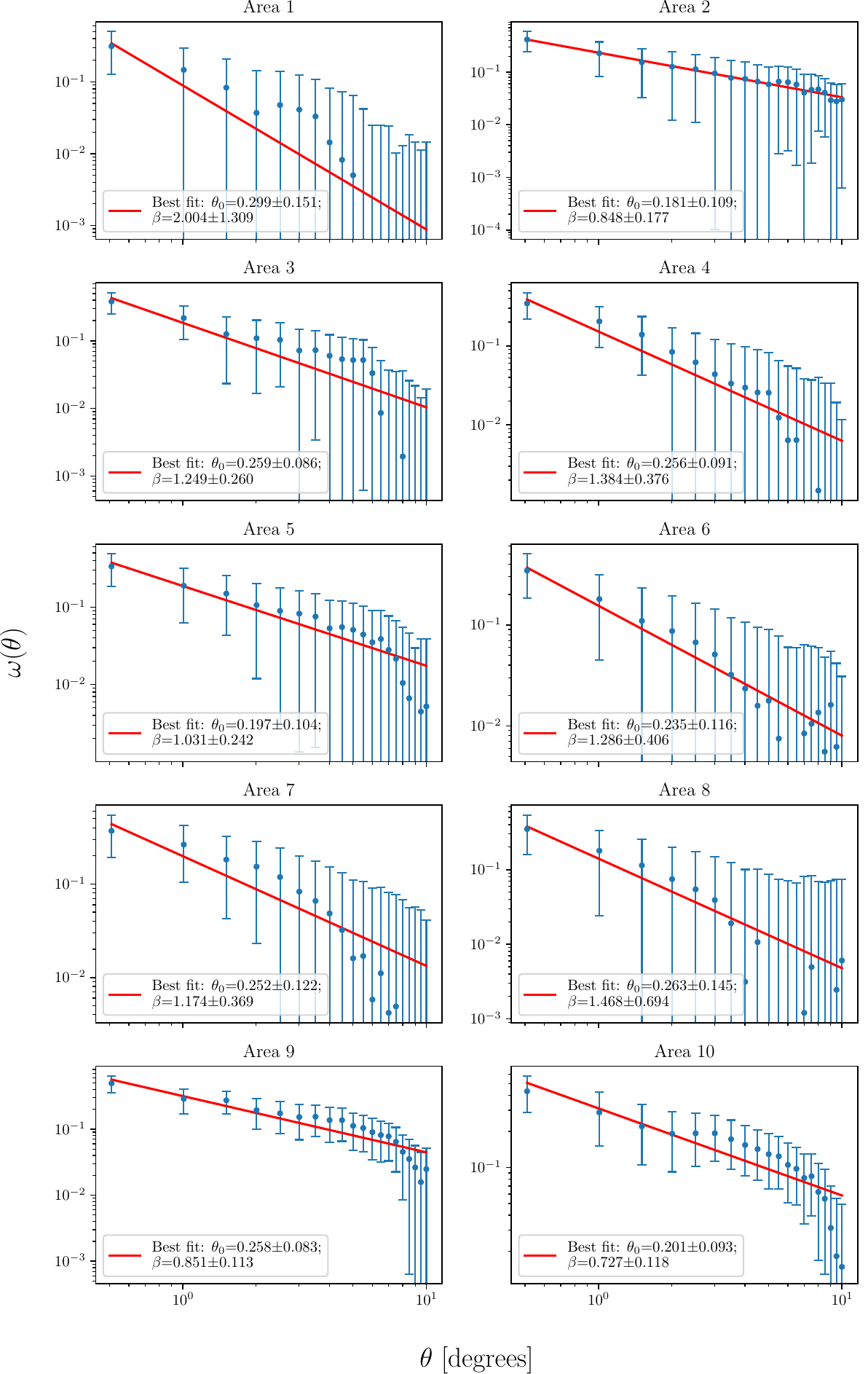}
\end{minipage}
\caption{Mosaic of the 2PACF for small angle analyses: angular 
distribution study of the 10 sky patches in which we divided the 
ALFALFA footprint, using the 2PACF with the LS estimator for a small 
angular interval, i.e., $0^{\circ} <\theta < 10^{\circ}$. 
}
\label{fig:tpacf-mosaic-loglog}
\end{figure*}

\begin{figure*}
\begin{minipage}[b]{\linewidth}
\includegraphics[width=8.4cm,height=6.5cm]
{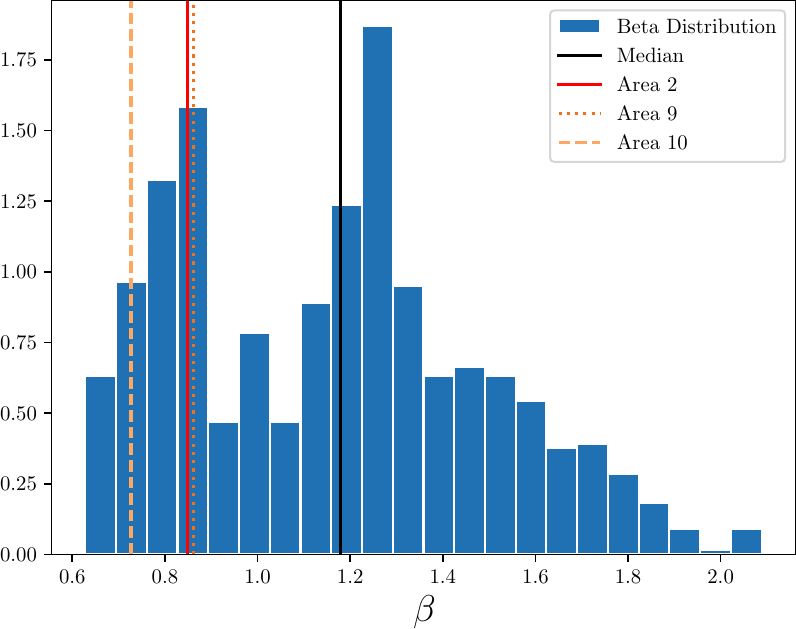}
\hspace{0.8cm}
\includegraphics[width=8.4cm,height=6.5cm]{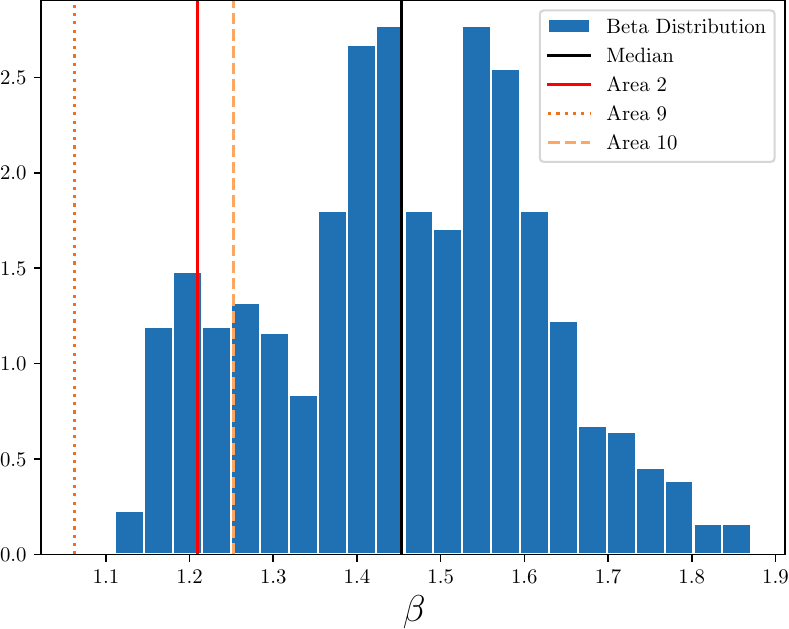}
\end{minipage}
\vspace{-0.3cm}
\caption{{\bf Left panel:} 
Distribution of the best-fit parameter 
$\beta^{S}$ from small-angle analyses of $1000$ {\bf Area}-mocks, the median and standard deviation are\, $\beta^{S;\,SI} = 1.180 \pm 0.325$, 
meaning that the 10 sky regions analysed are compatible with the hypothesis of statistical isotropy. 
The values for the {\bf Areas 2, 9,} and {\bf 10} are plotted as continuous, dotted, and dashed vertical lines, respectively (their corresponding $\beta$ values can be seen in Table~\ref{tab:best-fit}). 
All the $\beta^{L}$ parameters from the ALFALFA regions are within 
$2 \sigma$ CL. 
{\bf Right panel:} Distribution of the best-fit parameter 
$\beta^{L}$ from large-angle analyses of $1000$ {\bf Area}-mocks, the median and standard deviation are\, $\beta^{L;\,SI} = 1.453 \pm 0.161$.
The values for the {\bf Areas 2, 9,} and {\bf 10} are plotted as continuous, dotted, and dashed vertical lines, respectively (their corresponding $\beta^{L}$ values can be seen in Table~\ref{tab:best-fit}). 
Notice that the $\beta^{L}$ value for {\bf Area 9} appears with slightly higher statistical significance $2.4\,\sigma$; 
all the other $\beta^{L}$ parameters from the ALFALFA regions are within 
$2 \sigma$ CL.
}
\label{fig:beta-dist}
\end{figure*}

\begin{figure*}
\begin{minipage}[b]{\linewidth}
\includegraphics[width=8.4cm,height=6.5cm]
{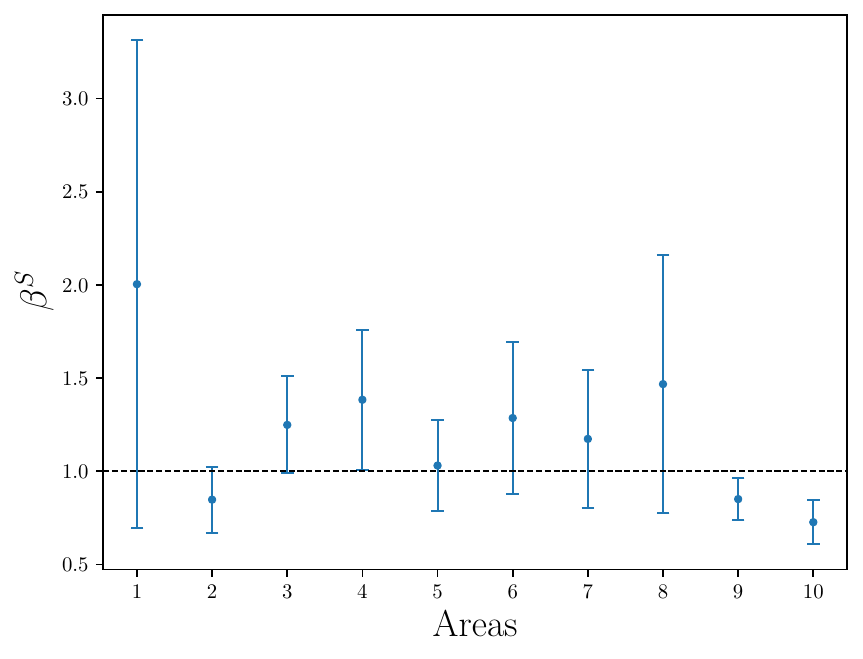}
\hspace{0.8cm}
\includegraphics[width=8.4cm,height=6.5cm]{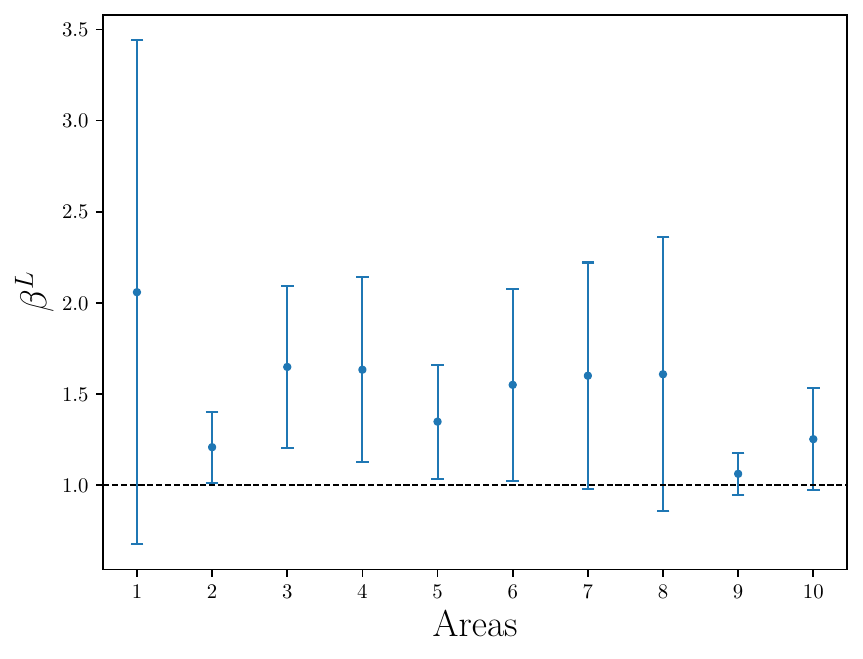}
\end{minipage}
\vspace{-0.3cm}
\caption{In these plots we show the 
values $\beta^{S}$ (left panel) and $\beta^{L}$ (right panel) obtained for the 10 {\bf Areas} in analysis (see Table~\ref{tab:best-fit}). 
As a reference, we draw a dashed line at \,$\beta = 1$.
}
\label{fig:beta-area}
\end{figure*}

\begin{figure*}
\begin{minipage}[b]{\linewidth}
\includegraphics[width=8.4cm,height=6.5cm]
{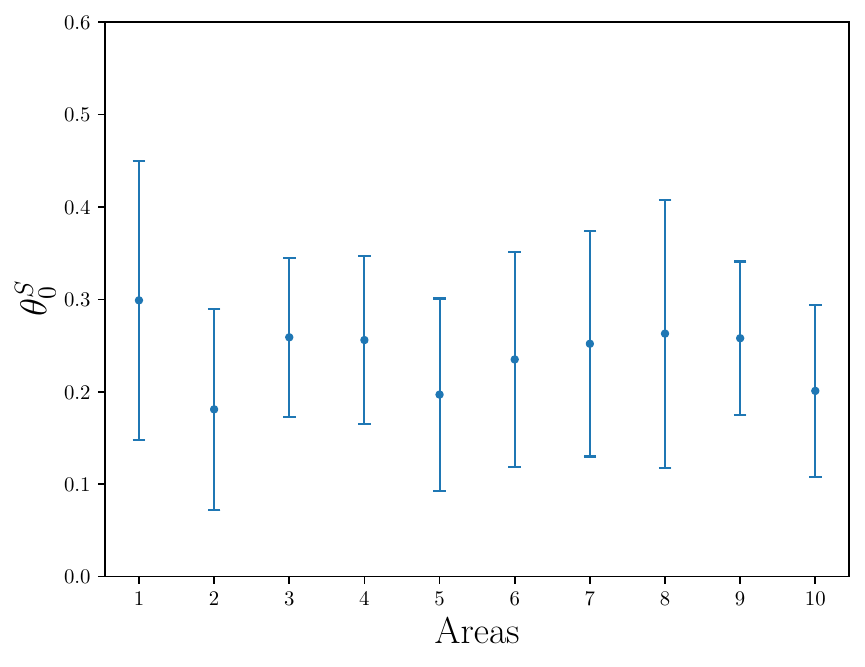}
\hspace{0.8cm}
\includegraphics[width=8.4cm,height=6.5cm]{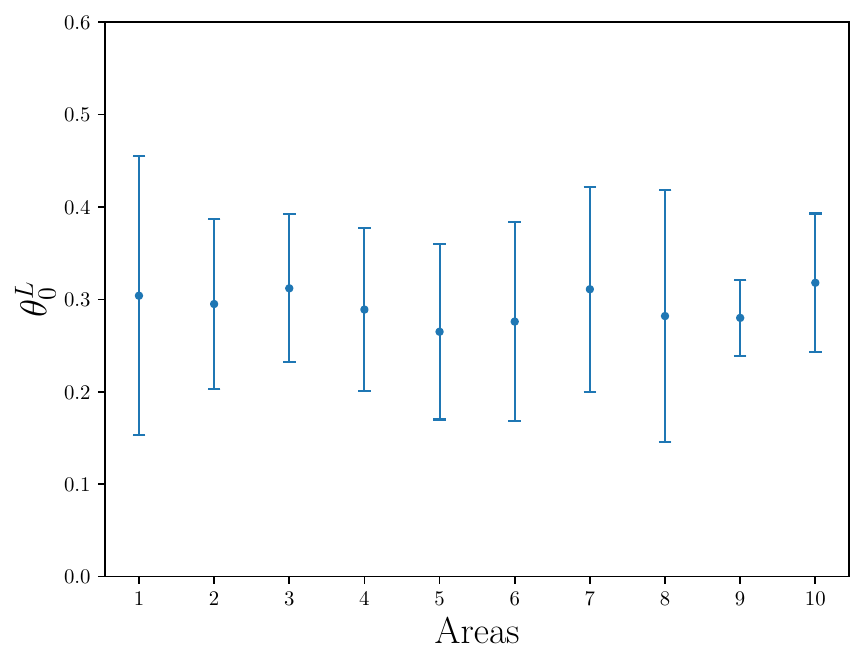}
\end{minipage}
\vspace{-0.3cm}
\caption{In these plots we show the 
values $\theta_0^{S}$ (left panel) and $\theta_0^{L}$ (right panel) obtained for the 10 {\bf Areas} in analysis (see Table~\ref{tab:best-fit}).}
\label{fig:theta-area}
\end{figure*}

The summary of our 2PACF small-angle analyses can be seen in Figure~\ref{fig:tpacf-mosaic-loglog} and in Table~\ref{tab:best-fit}, and complemented by the left panels of Figures~\ref{fig:beta-dist},~\ref{fig:beta-area},~and~\ref{fig:theta-area}.

\begin{table*}
\centering
\caption{Best-fit parameters considering the power-law relationship: 
$\omega(\theta) = (\theta / \theta_0)^{-\beta}$. 
We have performed two best-fit analyses, for {\em small angles} 
(i.e., $0^{\circ} <\theta < 10^{\circ}$) and for {\em large angles} 
(i.e., $0^{\circ} <\theta < 30^{\circ}$); we differentiate these 
cases with the super-index letter $S$ or $L$.  
}
\scalebox{1.25}{
\begin{tabular}[t]{lcccc}
\hline
& $\theta_0^{S}$ [degrees] & $\beta^{S}$ & $\theta_0^{L}$ [degrees] & $\beta^{L}$ \\
\hline
Area 1&  $0.299\,\pm\,0.151$ & $2.004\,\pm\,1.309$ & $0.304\,\pm\,0.151$ & $2.059\,\pm\,1.382$\\
Area 2&  $0.181\,\pm\,0.109$ & $0.848\,\pm\,0.177$ & $0.295\,\pm\,0.092$ & $1.209\,\pm\,0.195$\\
Area 3&  $0.259\,\pm\,0.086$ & $1.249\,\pm\,0.260$ & $0.312\,\pm\,0.080$ & $1.649\,\pm\,0.446$\\
Area 4&  $0.256\,\pm\,0.091$ & $1.384\,\pm\,0.376$ & $0.289\,\pm\,0.088$ & $1.634\,\pm\,0.506$\\
Area 5&  $0.197\,\pm\,0.104$ & $1.031\,\pm\,0.242$ & $0.265\,\pm\,0.095$ & $1.349\,\pm\,0.312$\\
Area 6&  $0.235\,\pm\,0.116$ & $1.286\,\pm\,0.406$ & $0.276\,\pm\,0.108$ & $1.551\,\pm\,0.525$\\
Area 7&  $0.252\,\pm\,0.122$ & $1.174\,\pm\,0.369$ & $0.311\,\pm\,0.111$ & $1.601\,\pm\,0.621$\\
Area 8&  $0.263\,\pm\,0.145$ & $1.468\,\pm\,0.694$ & $0.282\,\pm\,0.136$ & $1.609\,\pm\,0.751$\\
Area 9&  $0.258\,\pm\,0.083$ & $0.851\,\pm\,0.113$ & $0.280\,\pm\,0.041$ & $1.063\,\pm\,0.114$\\
Area 10& $0.201\,\pm\,0.093$ & $0.727\,\pm\,0.118$ & $0.318\,\pm\,0.075$ & $1.253\,\pm\,0.280$\\
\hline
\end{tabular}
}
\label{tab:best-fit}
\end{table*}

\subsection{2PACF large-angle analyses}

The best-fit parameter analyses for large angles $0^{\circ} < \theta < 30^{\circ}$, provides a measure of matter clustering at large angles in the 10 sky directions in which we divide the ALFALFA footprint. 
Then we compare them to the results obtained from similar analyses done in $1000$ {\bf Area}-mocks produced under the homogeneity and isotropy hypotheses. 
The best-fit parameter analyses performed on the $1000$ {\bf Area}-mocks, generated under the homogeneity and isotropy hypotheses, provide a distribution of $1000$ $\beta$ values, shown in the right panel of Figure~\ref{fig:beta-dist}, where the median and standard deviation of the best-fit parameter $\beta^{L}$ are \,$\beta^{L;\,SI} = 1.453 \pm 0.161$; the best-fit parameters for the ALFALFA regions is displayed in Table~\ref{tab:best-fit}. 
This comparison allows to reveal possible deviations of the best-fit parameters from the 10 regions in study with respect to the isotropic mocks. 
Our analyses let us to conclude that the 10 ALFALFA regions are compatible with the hypothesis of statistical isotropy within $<\!2\,\sigma$ CL, 
with the only exception of one region --{\bf Area 9}, located near the Dipole Repeller-- which appears slightly outlier ($2.4\,\sigma$).

\begin{figure*}
\begin{minipage}{\linewidth}
\centering
\includegraphics[width=15.0cm,height=22.3cm]{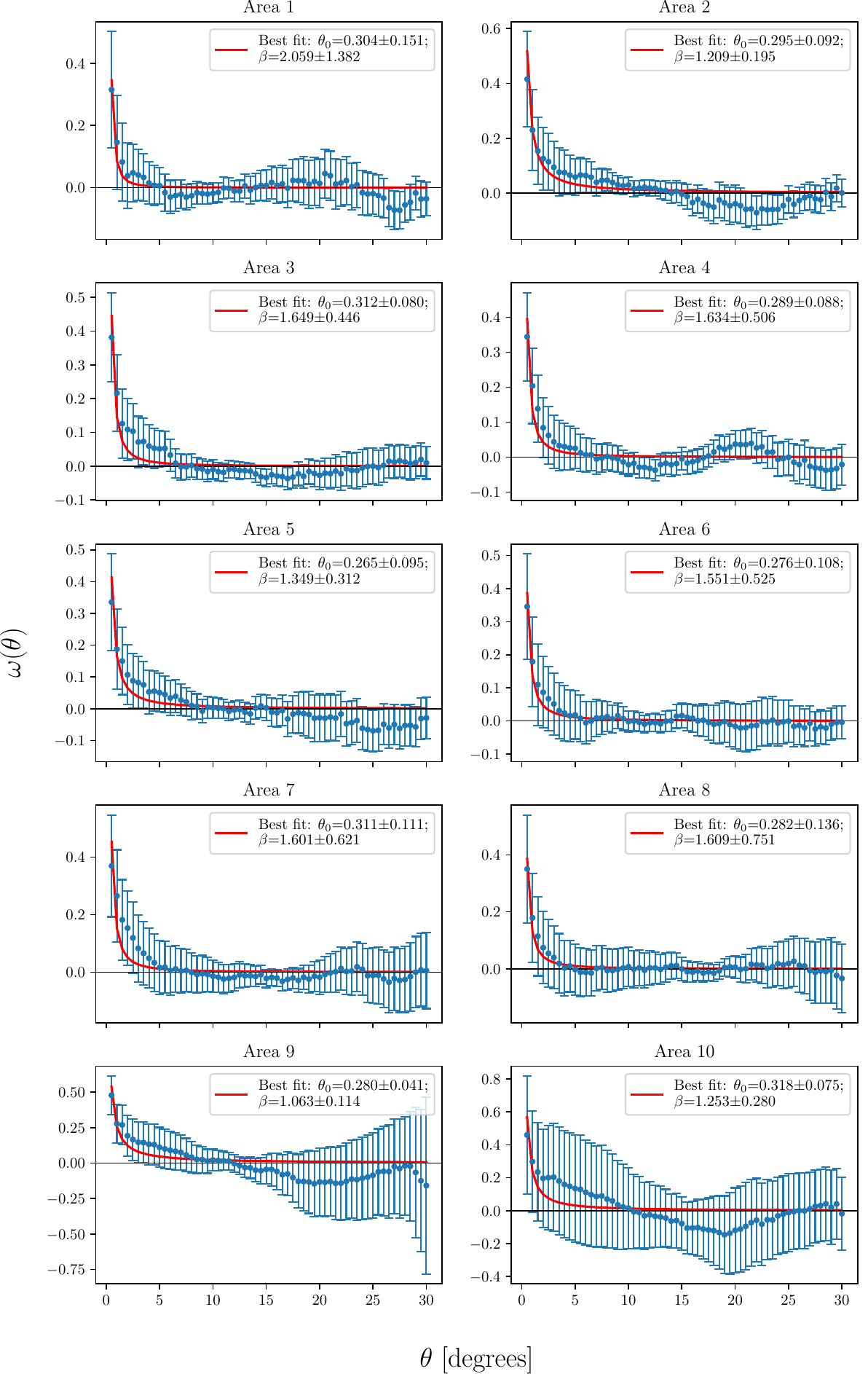}
\end{minipage}
\caption{Mosaic of the 2PACF for {\em large-angle analysis}: 
angular distribution study of the 10 sky patches in which we divided 
the ALFALFA footprint, using the 2PACF with the LS estimator for a 
large angular interval, i.e., $0^{\circ} <\theta < 30^{\circ}$. 
The comparison of the 2PACF from these 10 regions suggests further analyses that we perform in the section~\ref{intriguingfeatures}.
As discussed in the text, {\bf Areas 2, 9,} and {\bf 10} show, indeed, the prevalence of 3D large voids projected, partially, in these regions.}
\label{fig:tpacf-mosaic}
\end{figure*}

The summary of our 2PACF large-angle analyses, $0^{\circ} < \theta < 30^{\circ}$, can be seen in Figure~\ref{fig:tpacf-mosaic} and in Table~\ref{tab:best-fit}, and complemented by the right panels of Figures~\ref{fig:beta-dist},~\ref{fig:beta-area},~and~\ref{fig:theta-area}.

Interestingly, some regions show a dissimilar 2PACF at large-angles, 
in this sense the 2PACF provide complementary information to the best-fit parameter analyses.  
Observing the Figure~\ref{fig:tpacf-mosaic} one notices that the 2PACF 
corresponding to some regions present a consistent set of data points that evidence anti-correlations, manifested in the form of large deep depressions in the 2PACF, intriguing signatures that motivate us to investigate what could be causing them.

According to the literature the {\bf Area 2}, but also {\bf Areas 9} and {\bf 10}, are located close to regions where known underdense structures were reported~\citep{Tully87,Keenan,Moorman14,Hoffman17}. 
This turn these regions interesting arena to assess, with diverse statistical methodologies, to look for imprints left by these underdense structures in our analyses, and --possibly-- confirm their presence there.

\subsubsection{Cumulative Distribution Function}\label{CDF}

Motivated by the possibility of recognizing the presence of large underdense regions in some {\bf Areas} in study, in this section we use the Cumulative Distribution Function (CDF) to investigate if the features found in {\bf Area 2} could correspond to a projection of a set of small voids along the line-of-sight, as suggested by the analyses of~\cite{Moorman14}. 
For this purpose, we will use the distance information available in the ALFALFA catalogue; we represent distances by $r$. 
A similar analysis will be done studying also the {\bf Areas 9} and {\bf 10}.

The CDF is defined as the probability that a random variable $X$ has a value less than or equal to $x$ in the interval $F_X: \rm I\!R \rightarrow [0, 1]$~\citep{Wasserman04},
\begin{equation}
F_X(x) = \rm I\!P(X \leqslant x) \,.
\end{equation}
The theoretical curve is calculated for a Gaussian distribution with probability density function (PDF) defined as 
\begin{equation}\label{eq:gaussian}
P(x) = \frac{1}{\sigma \sqrt{2\pi}}\exp{\left[-\frac{1}{2} \left(\frac{x - \mu}{\sigma} \right)^{2} \right]} \,,
\end{equation}
where $\mu$, the mean, and $\sigma$, the standard deviation, 
are calculated from the ALFALFA distance data set of the 
{\bf Area} in analysis~\citep{Bevington03}.

\begin{figure}
\begin{minipage}[b]{\linewidth}
\centering
\includegraphics[scale=0.5]{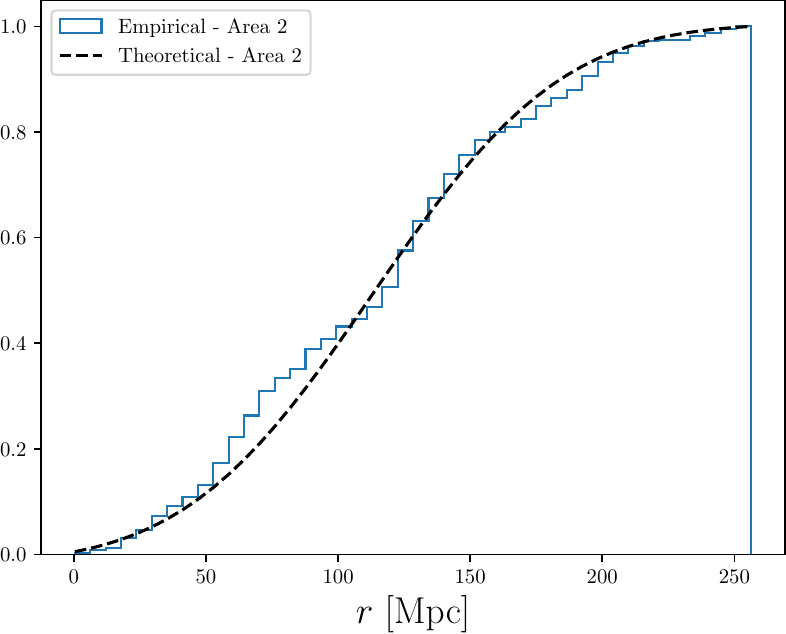}
\includegraphics[scale=0.5]{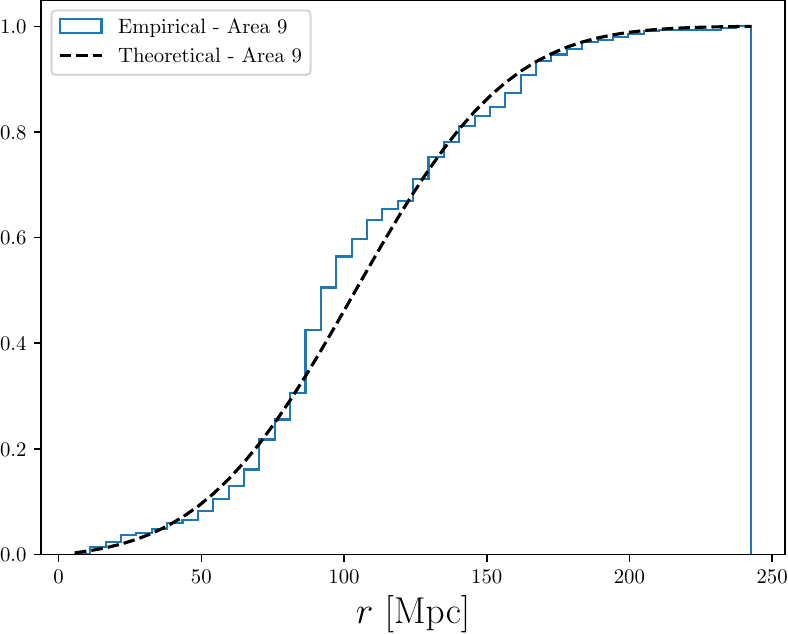}
\includegraphics[scale=0.5]{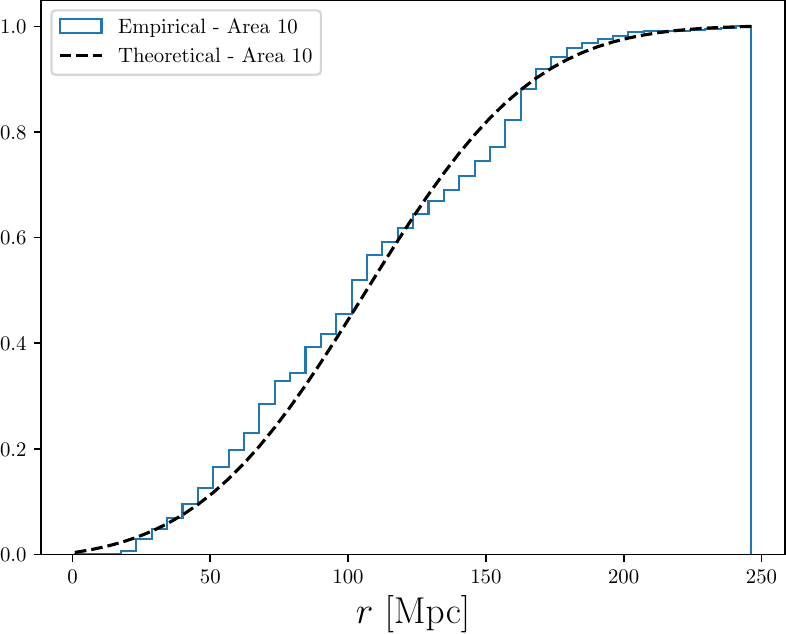}
\end{minipage}
\caption{Analyses of the {\bf Areas 2, 9}, and {\bf 10} with the Cumulative Distribution Function. 
{\bf Top Panel, Area 2:} This function shows clearly two depressions suggestive of underdense regions: one not so large around $r \in [100, 140]$ Mpc and the other is larger and deeper at $r \in [160, 245]$ Mpc. Due to the proximity of these depressions, and moreover, as suggested by the analyses of~\citet{Moorman14}, the voids network may interconnect small 
and medium size voids that can be interpreted as a very large void. 
Therefore, we conclude that the void structure projected onto {\bf Area~2} 
has an approximate length of $100\,\text{Mpc} \lesssim r \lesssim 245\,\text{Mpc}$, 
which corresponds to an underdense region with size around $150$ Mpc, 
and width of the order of $60$ Mpc.  
{\bf Middle Panel, Area 9:} Similarly, there is one depression around 
$r \in [40, 90]$ Mpc, and another around $r \in [140, 190]$ Mpc. 
{\bf Bottom Panel, Area 10:} This plot shows one deep depression around $r \in [130, 180]$ Mpc. 
In all cases the theoretical CDF was plotted using 44 bins, with mean $\mu = 116$ Mpc and standard deviation $\sigma = 56$ Mpc.}
\label{fig:cdf}
\end{figure}

We plot the theoretical CDF considering the basic features of the data in analysis and compute, and plot together, the CDF for the {\bf Area 2} data set (top panel of Figure~\ref{fig:cdf}). 
The theoretical CDF was plotted using 44 bins, with mean $\mu = 116$~Mpc and standard deviation 
$\sigma = 56$~Mpc. 
Examining the both CDFs, from {\bf Area 2} and the theoretical one, 
shown in Figure~\ref{fig:cdf} one clearly observes two depressions suggestive of underdense regions: 
one not so large around $r \in [100, 140]$ Mpc and the other is larger 
and deeper at $r \in [160, 245]$ Mpc. Due to the proximity of these depressions, and moreover, as suggested 
by the work of~\cite{Moorman14}, the voids network may interconnect small 
and medium size voids that can be interpreted as a very large void. 
Therefore, we conclude that the void structure projected onto {\bf Area 2} has the approximate location 
$100\,\text{Mpc} \lesssim r \lesssim 245\,\text{Mpc}$, 
which corresponds to an underdense region around $150$ Mpc in lenght, and width of the order of 
$60$~Mpc~\citep{Tully08,Tully13,Tully19}.

Our results are compatible with the analyses done by~\cite{Moorman14}, who studied the underdensity regions of ALFALFA in more detail (c.f. Figure 1 therein). 
The CDF of {\bf Area 2} shows the presence of distance intervals where there are fewer objects than expected when comparing with the Gaussian curve of Equation~(\ref{eq:gaussian}) 
and that, in fact, it would not correspond to a single giant void, but to some contiguous and (probably) connected smaller voids. 

Similar information can be deduced from the middle and bottom panel in 
Figure~\ref{fig:cdf}, analyses corresponding to the {\bf Areas 9} and {\bf 10}. 
In fact, in these plots we also observe the signature of considerable underdense regions.

\subsubsection{One, Two,... Three sky patches with large-angle features}
\label{intriguingfeatures}

According to our analyses of section~\ref{CDF} and  Appendix~\ref{ap:simulated-void}, the most notable large-angle features related to {\bf Areas 2, 9,} and {\bf 10} that make them peculiar regions are the following.

\begin{itemize}
\item 
Firstly, in Appendix~\ref{ap:simulated-void} we investigate the effect caused on the 2PACF when a sky region\footnote{The case studied in Appendix~\ref{ap:simulated-void} has the observational features of {\bf Area 2}.} 
contains a simulated void, i.e., an underdense region with number-density contrast $\delta < 0$\,\footnote{Actually, we performed several tests varying $\delta$ and the radius of the void; ultimately, the case that resulted similar to the 2PACF of {\bf Area 2} was obtained by simulating a void with $\delta = - 0.7$ and radius $9^{\circ}$.}: 
the result is that the 2PACF exhibits a consistent set of data points showing anti-correlations over a large angular range, although error bars make these data statistically consistent with $\omega(\theta)=0$. 
In other words, a simulated underdense region in a sky patch like {\bf Area 2} with number-density contrast $\delta = - 0.7$ produces a noticeable large and deep depression in the 2PACF, a peculiar behaviour --a {\em signature}, for short-- that is observed in the 2PACF of {\bf Areas 2, 9,} and {\bf 10} plotted 
in Figure~\ref{fig:tpacf-mosaic}. 
This result suggests, but does not prove, that the signature 
present in the 2PACF of {\bf Areas 2, 9,} and {\bf 10} 
could be associated with cosmic underdensities\footnote{ 
An {\em underdense} region, also referred as a {\em low-density} region or {\em void}, is a spatial region characterized by a density contrast $-1 \le \delta < 0$, which means less matter content inside that volume with respect to the density measured in a larger volume.}. 
\item
As a matter of fact, and widely reported in the literature, 
the {\bf Area 2} overlaps: 
(i) the LCV~\citep{Tully87,Tully19}; 
(ii) a huge low-density region~\citep{Keenan}; 
(iii) a 3D collection of medium-sized voids that projected 
on the celestial sphere appears as a large underdense region~\citep{Moorman14}. 
Moreover, we notice that the {\bf Areas 9} and {\bf 10} are close to the sky location where a large underdense region, termed the {\it Dipole Repeller}, was recently reported~\citep{Hoffman17}. 
\item
The Cumulative Distribution Function corresponding to the {\bf Areas 2, 9,} and {\bf 10}, discussed in detail in section~\ref{CDF}, provides interesting complementary information regarding the 3D geometry of the underdense regions we are investigating. 
The plots appearing in Figure~\ref{fig:cdf}, in particular the one corresponding to {\bf Area 2}, can be compared to the analyses done by~\cite{Moorman14} to confirm that the underdense regions reported there are indeed revealed in our data set analyses. 
\end{itemize}

\begin{figure}
\begin{minipage}[b]{\linewidth}
\centering
\mbox{\hspace{-0.4cm}
\includegraphics[width=8.9cm,height=7.4cm]{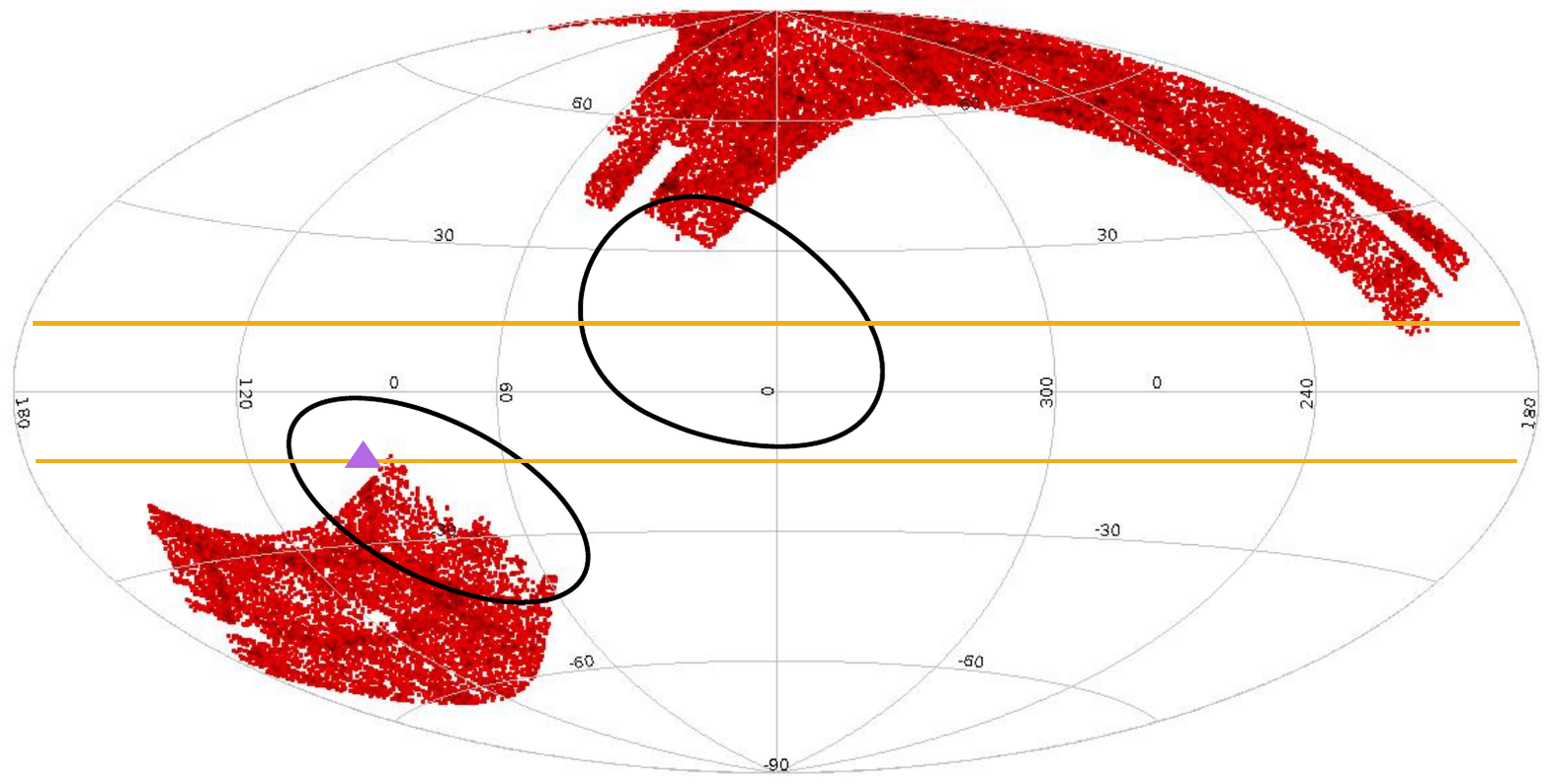}
}
\end{minipage}

\vspace{-0.8cm}
\caption{The ALFALFA footprint in galactic coordinates. 
The enclosed lines show, illustratively, the regions containing the projected underdense regions that could have left imprints in the data sample analysed. 
The horizontal lines, $\sim 15^{\circ}$ above and below the equator, represent the zone of avoidance, that is, the sky region where visible light is obscured by the Milky Way plane. The center of the picture corresponds to the galactic centre, and above and below the equator one has the NGH and SGH, respectively. The small (violet) triangle denotes the approximate position of the Dipole Repeller~\citep{Hoffman17}.
}
\label{mapaLCV-DR}
\end{figure}

All this information and analyses together confirms that {\bf Areas 2, 9,} and {\bf 10}\, contain the projection of cosmic voids, that is 3D large underdense regions that leave their signatures in our 2PACF analyses.  
The Figure~\ref{mapaLCV-DR} summarize these results illustrating the approximate positions, but not the true extensions, of the LCV and Dipole Repeller structures near the {\bf Areas 2, 9,} and~{\bf 10}.

\section{Conclusions and Final Remarks}\label{sec:final}

This work comprises model-independent analyses to study the statistical isotropy 
in the Local Universe exploring the astronomical data of the ALFALFA survey 
($0 < z < 0.06$). 
The enormous importance of the Cosmological Principle in the process of achieving 
a successful concordance cosmological model is recognized by the community. 
For this, it dedicates efforts to probe its validity, and search for possible restrictions~\citep{Marques18,Bengaly19,Sarkar2019,
Avila2023,Khan23}.

Although closer to us, the Local Universe is difficult to be observed in detail just because many 
structures lies behind the Milky Way disc, a structure that is opaque to visible electromagnetic wavelengths. 
Despite these inconveniences, there is still a large area mapped by astronomical surveys, 
with observed regions in opposite galactic hemispheres that can be investigated. 
In this work we complement previous homogeneity and isotropy studies by investigating in detail 
the angular distribution of the HI cosmological tracer, from the ALFALFA survey, which contains 
valuable information of the matter and voids structures in the Local 
Universe~\citep{Courtois13,Moorman14,Hoffman17,Tully19}. 
For this study we divided the ALFALFA footprint in a set of 10 patches as shown in Figure~\ref{fig:alfalfa-division}, 
with area $\sim \!\!700\,\text{deg}^{2}$ each. 
The construction of these 10 regions was a compromise because they cannot be so large that we have few regions 
for statistical analysis, and they can not be so small, with low number density in each region, 
such that its size be below the 2D homogeneity scale 
(for the ALFALFA catalogue this scale is $\theta_H \simeq 16^{\circ}$~\citep{Avila18}; more generally for matter 
with bias $b \simeq 1$ one has $\theta_H \simeq 20^{\circ}$~\citep{Avila19}). 
Then we use the 2PACF, with the LS estimator, to investigate the behavior of the angular distribution of the ALFALFA cosmic objects in each region, 
quantifying the behavior of the 2PACF through a power-law best-fit analyses, 
and determine the statistical significance of the best-fit parameters from the 10 ALFALFA regions by comparison with the parameters obtained through the same 
directional analyses procedure applied to a set of lognormal mock catalogues produced under the hypotheses of statistical homogeneity and isotropy. 
According to the results, summarized in Table~\ref{tab:best-fit} and complemented with the small- and large-angle analyses displayed in the histogram plots in Figure~\ref{fig:beta-dist}, our final conclusion is that the Local Universe, as mapped by the HI extragalactic sources of the ALFALFA survey, is in concordance with the hypothesis of statistical isotropy within $2\,\sigma$ CL, for small and large angles, with the only exception of the {\bf Area 9} \,--located near the Dipole Repeller-- which appears slightly outlier (2.4 $\sigma$).

Nonetheless, interesting features have been detected in the 2PACF large-angle analyses of the 10 ALFALFA regions studied. 
In fact, in Figure~\ref{fig:tpacf-mosaic} one notices that {\bf Areas 2, 9}, and {\bf 10} show a notable depression in their 2PACF 
compatible with a large underdensity, or with a collection of medium-size underdensities, which can have their origin in a cosmic void region with $\delta \simeq -0.7$.
Thanks to recent efforts dedicated to investigate the structures in the Local Universe, today we know about the spatial distribution 
of clustered matter and voids; in particular the three sky patches --{\bf Areas 2, 9,} and {\bf 10}-- partially coincide with known underdense regions, 
namely the LCV~\citep{Tully19} and the Dipole Repeller~\citep{Hoffman17}. 
The Figure~\ref{mapaLCV-DR} illustrates the approximate position, but not the true extension, of these underdensity regions, namely the LCV and the Dipole Repeller. 
Due to this valuable information, we have a reasonable explanation of the intriguing behavior observed in the 2PACF.

\section*{Acknowledgements}
CF and AB thank Coordenação de Aperfeiçoamento de Pessoal de Nível Superior (CAPES) and Conselho Nacional de Desenvolvimento Científico e Tecnológico (CNPq) for their grants under which this work was carried out. 
FA thanks CNPq and Fundação Carlos Chagas Filho de Amparo à Pesquisa do Estado do Rio de Janeiro (FAPERJ), Processo SEI 260003/014913/2023 for financial support.

\section*{Data Availability}

The data underlying this article will be shared on reasonable request to the corresponding author.



\bibliographystyle{mnras}
\bibliography{refs} 



\appendix
\section{A simulated underdense region: a void}\label{ap:simulated-void}

In this appendix, we evaluate the performance of the 2PACF, using the 
Landy-Szalay~\citep{LS93} estimator, to reveal the signature that can be produced 
by a simulated void, or underdense region, in a sky patch similar to {\bf Area 2}. 


We perform an interesting consistency test, our objectives are: \\
(i) to confirm if an artificial underdense region leave a signature in the 2PACF; 
\,(ii) if yes, what is the size and the suitable number-density contrast, $\delta$, that leave a signature similar to that one observed in the 2PACF 
of {\bf Area 2}?

Consider a simulated sky region with the same features as the {\bf Area 2}, that is, same angular scale dimensions and number of cosmic objects therein; 
remove 30\% of objects located inside a disc of $9^{\circ}$ of radius\footnote{The location of the disc centre that simulates  the artificial void is not relevant because for the 2PACF what matters is the distance between pairs.} and redistribute these objects uniformly outside the disc. 

\begin{figure*}
\mbox{
\hspace{-0.1cm}
\includegraphics[width=5.8cm,height=6.3cm]{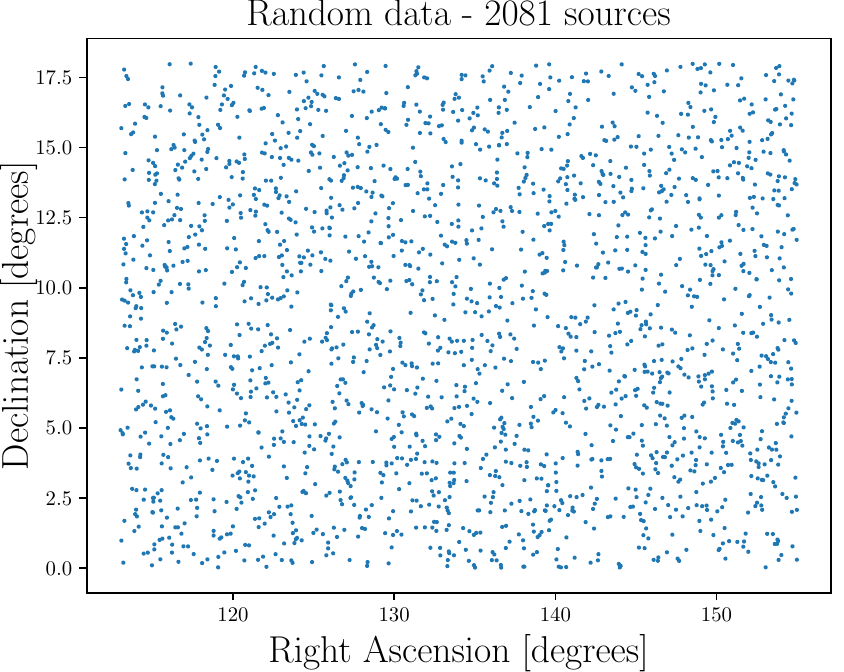}
\hspace{-0.1cm}
\includegraphics[width=5.8cm,height=6.3cm]{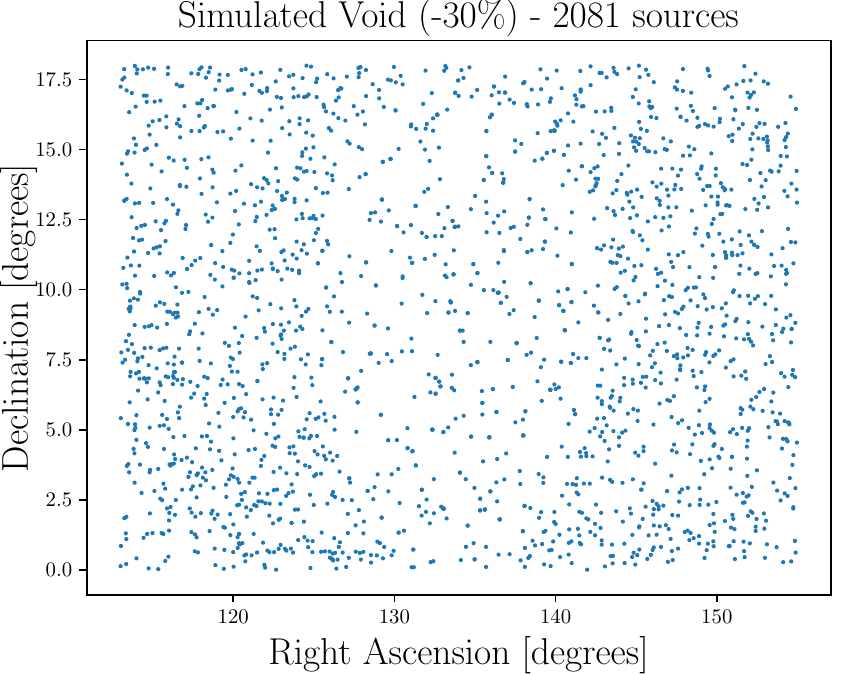}
\hspace{-0.2cm}
\includegraphics[width=7.0cm,height=6.8cm]{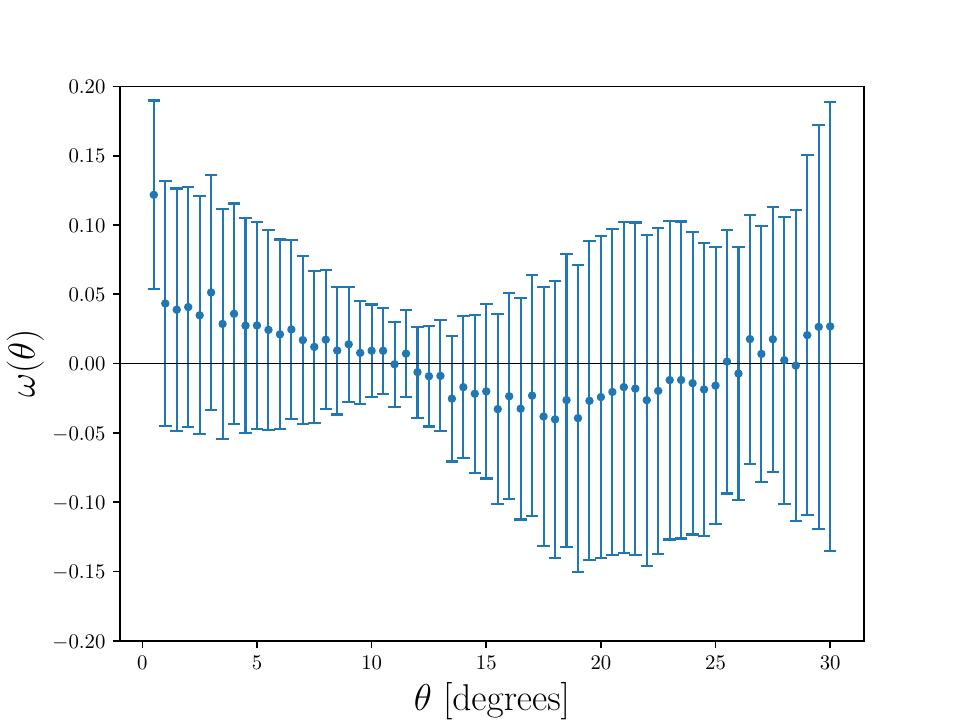}
}
\caption{Analysis of a simulated void region, with $\delta = -0.7$, for a region with the observational features of {\bf Area 2} (i.e., covered sky area, and number density of cosmic sources).  
{\bf Left panel:} 
Simulated region with 2081 uniformly distributed  objects. 
{\bf Middle panel:} 
The same region but with 30\% of the cosmic objects removed from a disc of $9^{\circ}$ and distributed randomly outside the disc. This means that the number density continues to be the same as in the left panel. 
{\bf Right~panel:} 
2PACF obtained with the LS estimator for the area shown in the middle panel, that is, containing a simulated void with $\delta = -0.7$ and radius of $9^{\circ}$.
} 
\label{fig:toy-model-void}
\end{figure*}

After performing several tests varying $\delta$ and the radius of the void, we find the case that resulted similar to the 2PACF of {\bf Area 2} was obtained by simulating a void with $\delta = - 0.7$ and radius $9^{\circ}$. 
Our results can be observed in Figure~\ref{fig:toy-model-void} where we show the original region with uniformly distributed data and the region after removing 30\% of the objects from a disc of $9^{\circ}$ radius. 
The 2PACF analysis of this region with a simulated void is done using the LS estimator and the result is shown in the right panel of 
Figure~\ref{fig:toy-model-void}. 
Importantly, although the error bars make the 2PACF compatible with zero angular correlations, this does not mean that the {\bf Area 2} corresponds to a uniform distribution of sources. 
This conclusion is due to the fact that the right panel was indeed produced by a simulated void with $\delta = -0.7$ and radius of $9^{\circ}$, which in cosmic terms means something extraordinary~\citep{Keenan}. 

From our results displayed in Figure~\ref{fig:toy-model-void}, we conclude that a void characterized with number density contrast $\delta = -0.7$ clearly leave a signature that can be detected with the 2PACF estimator. 

\begin{figure*}
\mbox{
\hspace{-0.1cm}
\includegraphics[width=5.9cm,height=5.9cm]{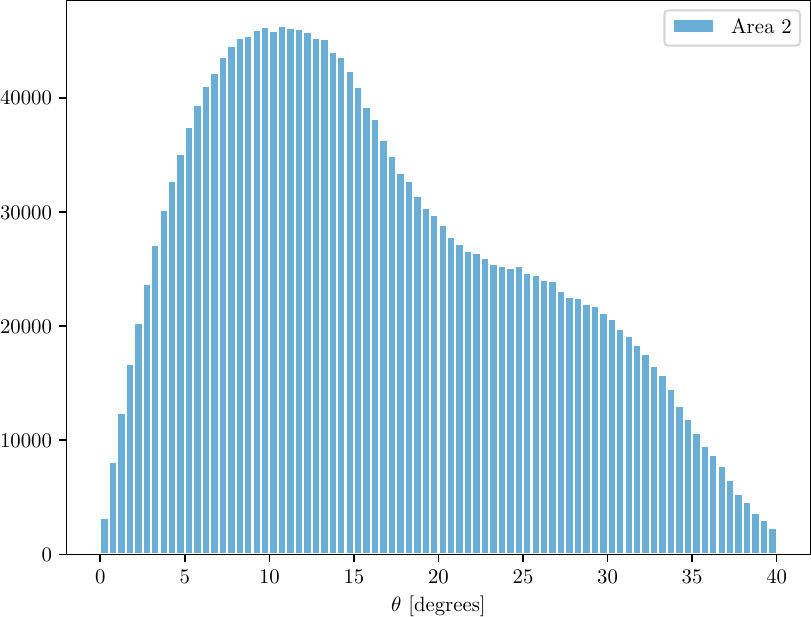}
\hspace{0.1cm}
\includegraphics[width=5.9cm,height=5.9cm]{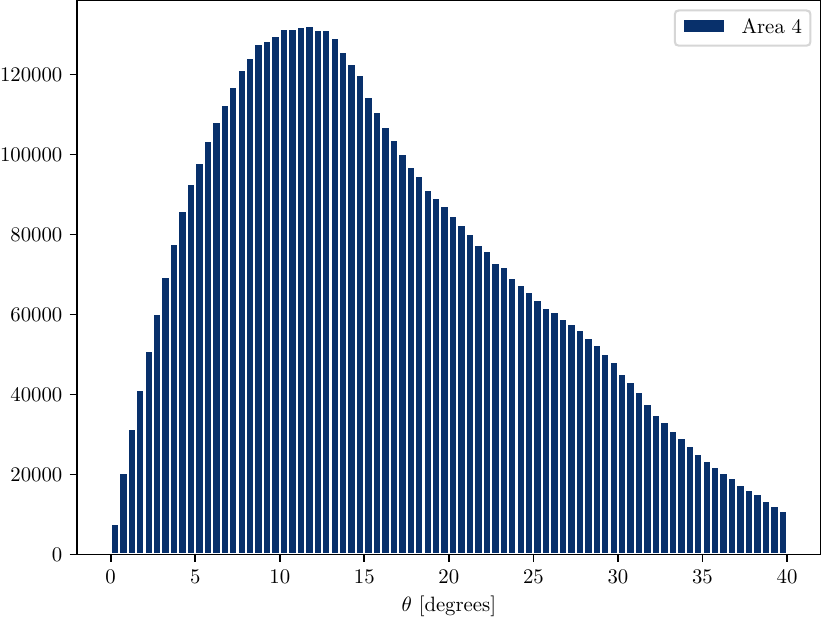}
\hspace{0.1cm}
\includegraphics[width=5.9cm,height=5.9cm]{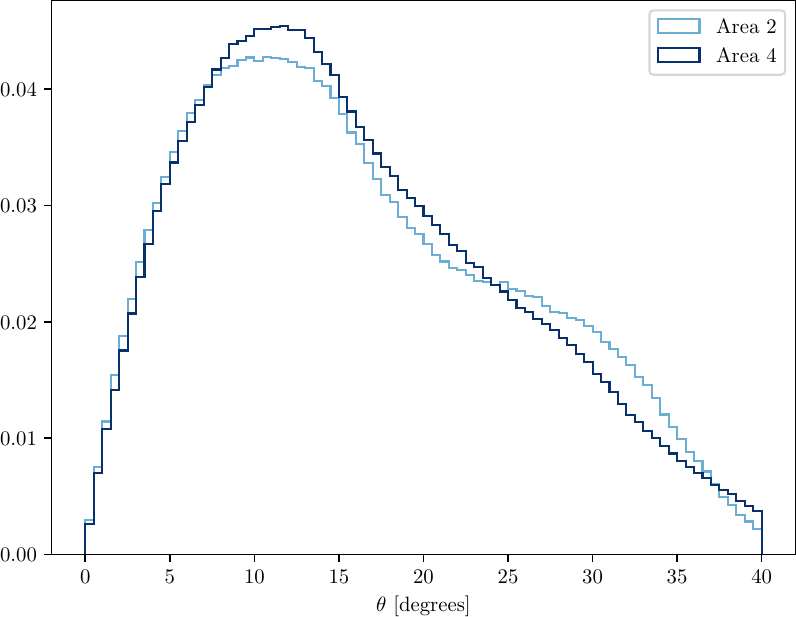}
}
\caption{{\bf Left panel:}
Histogram of the number of data pairs, as a function of the angular separation $\theta$, corresponding to {\bf Area 2}.
{\bf Middle panel:}
Histogram of the number of data pairs, as a function of the angular separation $\theta$, corresponding to {\bf Area 4}.
{\bf Right panel:}
Normalized histograms of the number of data pairs from 
{\bf Areas 2} and {\bf 4} to compare their behaviour as a function of $\theta$. 
As observed, {\bf Area 2} shows a large-angle interval of anti-correlations between pairs, aproximately at the same scales as those shown in the right panel of Figure~\ref{fig:toy-model-void}, namely $10^{\circ} \lesssim \theta \lesssim 25^{\circ}$. 
} 
\label{figsextras}
\end{figure*}

As a complementary test of this analysis, we perform a comparison between the number of distance pairs of cosmic objects in {\bf Areas 2} and {\bf 4}.
As observed in Figure~\ref{figsextras}, the {\bf Area 2} shows a substantial lack of distance pairs at the angular scales 
$10^{\circ} \lesssim \theta \lesssim 25^{\circ}$. 
All these analyses confirms our conclusion, mentioned in section~\ref{intriguingfeatures}, that the 2PACF indeed shows a signature corresponding to a large underdense region in 
{\bf Area 2}, and by extension, also in {\bf Areas 9} and~{\bf 10}. 

\section{Consistency check: a null test}\label{ap:tests}

\begin{figure}
\begin{minipage}{\linewidth}
\centering
\includegraphics[width=\textwidth]{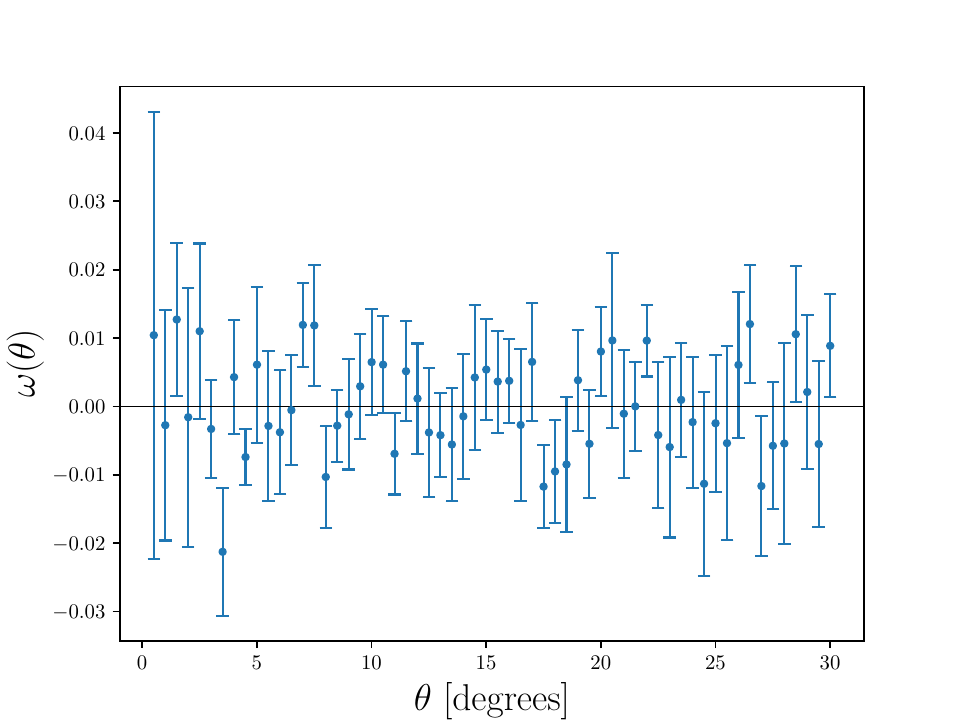}
\end{minipage}
\caption{Null test for the random data set. 
We perform the 2PACF analysis considering one random catalogue as a {\it pseudo}-data catalogue. 
As observed, the 2PACF shows no evidence of any signature, just statistical fluctuations, as expected.}
\label{fig:random-tpacf}
\end{figure}

To probe the statistical isotropy of the Local Universe we based our methodology of 
study in the 2PACF. 
To give support to our findings, we elaborate a consistency test, the so-called {\em null test} that helps to ensure that there is no signature present in our random catalogues 
that can bias our results. 
We shall test a random catalogue to verify that it does not produce any signature in the 2PACF. 
The result can be observed in Figure~\ref{fig:random-tpacf}, where we observe that the 2PACF shows no evidence of any signature, just statistical fluctuations, as expected. 

\bsp	
\label{lastpage}
\end{document}